\documentclass[12pt, prd, showpacs]{revtex4}
\usepackage{amssymb}
\usepackage{amsmath}
\usepackage{graphicx}

\begin{document}

\title{Generalized Lema\^{\i}tre time for rotating and charged black holes
and its near-horizon properties}
\author{A. V. Toporensky}
\affiliation{Centre for Cosmology and Science Popularization (CCSP),}
\affiliation{SGT University, Gurugram, Delhi-NCR, Haryana 122505, India}
\email{atoporensky@gmail.com}
\author{O. B. Zaslavskii}
\affiliation{Department of Physics and Technology, Kharkov V.N. Karazin National
University, 4 Svoboda Square, Kharkov 61022, Ukraine}
\email{zaslav@ukr.net }

\begin{abstract}
We consider the behavior of the analogue of the Lema\^{\i}tre time when a
particle approaches the horizon of a rotating black hole. For the Kerr
metric, the aforementioned time coincides with the Doran or Natario time but
we consider a more general class of metrics. We scrutiny relationship
between (i) its finiteness or divergence, (ii) the forward-in-time
condition, (iii) the sign of a generalized momentum/energy, (iv) the
validity of the principle of kinematic censorship. The latter notion means
impossibility to release in any event an energy which is literally infinite.
As a consequence, we obtain a new explanation, why collisions of two
particles inside the horizon do not lead to infinite energy in their center
of mass frame. The same results are also obtained for the Reissner-Nordstr%
\"{o}m metric.
\end{abstract}

\maketitle

\section{Introduction}

Exact solutions of field equations describing spherically symmetric black
holes were discovered in coordinates that are spoiled on the horizon, so one
is led to search for new coordinates that do not have \ this drawback. This
issue belongs to fundamentals of black hole physics and this line or
research has been continuing until now. Among these metrics, one of the
famous frames is the Lema\^{\i}tre one \cite{lem}, found for the
Schwarzschild metric \cite{LL} and admitting generalizations to other
spherically symmetric space-times. Meanwhile, the situation becomes more
difficult and subtle for rotating black holes. In the first place, this
concerns the Kerr metric. Originally, it was found \cite{kerr} in
coordinates that are horizon-penetrating and present it in a form that
resembles the Eddington-Filkenstein one for the Schwarzschild metric.
However, for practical purposes and for solving physical and astrophysical
problems, another coordinate frame is more popular - the Boyer-Lindquist one 
\cite{boyer}. Meanwhile, once it is adopted, the same difficulty as in the
Schwarzschild case arises again since these coordinates are spoiled on the
horizon. To repair this drawback, rather recently a form of the Kerr metric
based on a family of observers falling from infinity and crossing the
horizon was suggested for the Kerr metric in two versions \cite{doran}, \cite%
{natario}. Generalization of both approaches, valid for a generic stationary
axially symmetric metric describing a rotating black hole, has been
suggested in \cite{18}. By analogy with the Schwarzschild metric, we call a
corresponding time variable a Lema\^{\i}tre one. Let us recall that the
Schwarzschild metric one can also use so-called the Painlev\'{e}-Gullstrand
frame \cite{pain}, \cite{gull}. In both frames the time variable is the same
but the Lema\^{\i}tre form is diagonal in contrast to the Painlev\'{e}%
-Gullstrand frame. This is achieved due to an appropriate transformation of
a spatial coordinate in the Lema\^{\i}tre case. For rotating back holes, it
is also natural to use the same terminology and call "Lema\^{\i}tre time" a
time variable that is connected with observers falling from the rest (i.e.
with zero angular momentum and the Killing energy being equal to $mc^{2}$)
at infinity. If one manages to construct it, the Lema\^{\i}tre time actually
coincides with the synchronous time. This problem is rather difficult for
the Kerr metric \cite{khat} but we will use the aforementioned term anyway.

In the present paper we consider only realistic black holes. It is known
that already for Schwarzschild metric the Lema\^{\i}tre time does not cover
the whole geodesically complete space-time of an eternal vacuum black-white
hole, and there are four possible ways to define a Lema\^{\i}tre-like time
covering different parts of the whole picture \cite{radobs}. We will
restrict ourselves here by the classical way to construct the Lema\^{\i}tre
time. For an example of using another Lema\^{\i}tre-like time to describe
processes near a hypothetical realistic white hole, see \cite{white-black}.%
\textbf{\ }

The aim of the present work is to elucidate properties of the Lema\^{\i}tre
time when a particle trajectory approaches the horizon. Motivation here is
twofold. First, we find it necessary to describe the properties of the
aforementioned metrics as completely as possible and elucidate, how particle
motion looks in these frame in a physically interesting region - the
vicinity of the horizon. This belongs to a rather traditional line of
research, where properties of regular frames in black hole space-times
remain an "eternal" subject.

Second, there is more specific need for this task due to the fact that
during last decade a special attention was focused on near-horizon high
energy particle collisions. The starting point was discovery of the Ba\~{n}%
ados-Silk-West (BSW) effect \cite{ban}. According to this effect, the energy 
$E_{c.m.}$ in the center of mass of two particles that collide near the
horizon can, under some conditions, be unbounded from above. For such a
situation to be realized near the outer horizon (this is necessary if we
want this effect to be visible by a remote observer), one of particles
should be fine-tuned or near-fine-tuned. This line of research remains
active including astrophysical context (see, for example, recent works \cite%
{ms}, \cite{dgd},) However, for collisions near the inner horizon (if it
exists) this fine-tuning is not needed and conditions for collision energy
to be extremely high are much weaker. Though a remote observer has no access
to physical observations inside a black hole, such a situation is
interesting from a theoretical point of view. What makes it even more
drastic, is the paradox according to which the collision energy can be
literally infinite for collisions exactly at an inner horizon, which is
physically inappropriate.

Detailed discussion of this seeming contradiction was done in \cite{lake2}, 
\cite{grib1}, \cite{grib2}, \cite{inner}. The corresponding analysis relies
on description of the space-time structure. Meanwhile, the present work
suggests another explanation of this fact.

In \cite{cens} a more general concept of "kinematic censorship" has been put
forward. Typical resolution of the paradoxes of the type under discussion,
in agreement with this principle, consists in that event of collision does
not take place at all. In terms of the Lema\^{\i}tre time, this happens if
for one colliding particle it remains finite whereas for the second particle
it diverges near the horizon. Thus, there is crucial dependence between
possibility of high energy particle collisions (the BSW effect or its
analogue near the inner horizon) and behavior of the time variable under
discussion, when a point of observation approaches the horizon.

Therefore, the analysis of properties of the Lema\^{\i}tre time is required
for better understanding of kinematics and dynamics of high energy particle
collisions near the horizon, both for the event horizon and the inner one.

\section{Rotating black hole: general set-up\label{II}}

Let us start with a generic metric describing axially symmetric rotating
black hole:

\begin{equation}
ds^{2}=-N^{2}dt^{2}+g_{\phi }(d\phi -\omega dt)^{2}+\frac{dr^{2}}{A}%
+g_{\theta }d\theta ^{2}.  \label{met}
\end{equation}

The surface where $N=0$, $A=0$ corresponds to an outer $r=r_{+}$ or inner $%
r=r_{-}$ horizon. We assume that the metric coefficients do not depend on $t 
$ and $\phi $. Correspondingly, for a particle moving in this background
there exist integrals of motions. These are the energy $E$ and angular
momentum $L$. We also assume the symmetry of the metric with respect to the
equatorial plane $\theta =\frac{\pi }{2}$ and restrict ourselves by particle
motion just within this plane. Then, it follows from equations of motion
that for a particle moving freely%
\begin{equation}
m\dot{t}=\frac{X}{N^{2}}\,,
\end{equation}%
\begin{equation}
m\dot{\phi}=\frac{L}{g_{\phi }}+\frac{\omega X}{N^{2}}\text{,}
\end{equation}%
\begin{equation}
m\frac{\dot{r}}{\sqrt{A}}=\sigma \frac{P}{N}\text{,}
\end{equation}%
where%
\begin{equation}
X=E-\omega L\text{,}  \label{x}
\end{equation}%
\begin{equation}
P=\sqrt{X^{2}-N^{2}(m^{2}+\frac{L^{2}}{g_{\phi }})}\text{,}
\end{equation}%
point denotes a derivative with respect to the proper time $\tau $, and $%
\sigma =\pm 1$ depending on the direction of motion. Near a black hole
horizon, $\sigma =-1$ since a particle moves towards the horizon, whereas
near a white hole one $\sigma =+1$ since a particle moves away from it.
Outside the horizon the forward-in-time condition gives us%
\begin{equation}
X>0.  \label{forw}
\end{equation}

More precisely, on the horizon itself $X=0$ is also possible but we do not
consider such fine-tuned (critical) trajectories.

It follows from these equations that%
\begin{equation}
\frac{dt}{dr}=\sigma \frac{X}{\sqrt{A}PN}.
\end{equation}

\section{Dirty black hole: equatorial motion\label{dirty}}

We assume that%
\begin{equation}
N^{2}=\alpha \Delta \text{, }A=\frac{\Delta }{\rho ^{2}}\text{,}
\end{equation}%
where $\Delta =0$ on the horizon. Further, we describe briefly the procedure
that enables us to make the metric coefficients on the horizon finite and
nonzero. We follow \cite{18}, where a reader can find more detailed
derivation and discussion. Everywhere in formulas below, we put $\theta
=const=\frac{\pi }{2}.$

At the horizon $N=0$ and $A=0$, so the metric fails to be regular. To repair
this shortcoming, let us make the coordinate transformations 
\begin{equation}
dt=d\bar{t}-\frac{zdr}{\Delta }\text{,}  \label{t2}
\end{equation}%
\begin{equation}
d\phi =d\bar{\phi}+\frac{\xi dr}{\Delta }
\end{equation}%
where 
\begin{equation}
\xi -\omega z=h\Delta \text{,}
\end{equation}%
\begin{equation}
\mu =\frac{\rho ^{2}-\alpha ^{2}z}{\Delta }\text{,}
\end{equation}%
the functions $\mu $, $\xi $, $z$ and $h$ depending on $r$. Then, the metric
on the plane $\theta =\frac{\pi }{2}$ reads%
\begin{equation}
ds^{2}=-d\bar{t}^{2}\frac{\alpha \rho ^{2}}{\mu }+\mu (dr+\frac{\alpha z}{%
\mu }d\bar{t})^{2}+g_{\phi }(d\bar{\phi}-\omega d\bar{t}+hdr)^{2}\text{.}
\label{mum}
\end{equation}%
It generalizes the results \cite{doran}, \cite{natario} derived for the Kerr
metric. We require the functions $\mu $ and $h$ to remain regular on the
horizon. To this end, we choose them such that on the horizon 
\begin{equation}
z^{2}\alpha =\rho ^{2}  \label{eq}
\end{equation}%
with $z>0$.

\subsection{Particular case: simplified version of metric}

It is instructive to give more explicit form of the metric to trace its
analogy with the spherically symmetric one. Without the loss of generality,
we can consider the case $\mu =\alpha \rho ^{2}$ which simplifies formulas.
It is convenient to absorb $\rho $ by $\Delta $, so we put $\rho =1.$ As for
the metric (\ref{mum}), the angle $\theta $ is excluded, so one can make
redefinition of the radial coordinate $r\rightarrow \bar{r}=\sqrt{\alpha }r$%
. Also, one can put $h=0$. This is quite sufficient for our goal to make the
metric coefficients regular near the horizon. Then, 
\begin{equation}
ds^{2}=-d\bar{t}^{2}+(d\bar{r}+\bar{v}dt)^{2}+g_{\phi }(d\bar{\phi}-\omega d%
\bar{t})^{2}\text{,}
\end{equation}%
where $\bar{v}=z\sqrt{\alpha }$. This looks like a rotational version of the
Painlev\'{e}-Gullstrand \cite{pain}, \cite{gull} coordinate system. One can
make further transformation of the spatial coordinate retaining the same $%
\bar{t}$ to obtain the rotational version of the Lema\^{\i}tre system:%
\begin{equation}
ds^{2}=-d\bar{t}^{2}+\bar{v}^{2}d\chi ^{2}+g_{\phi }(d\bar{\phi}-\omega d%
\bar{t})^{2},
\end{equation}%
where 
\begin{equation}
\chi =\bar{t}+\int \frac{d\bar{r}}{(1-\bar{v}^{2})\bar{v}}.
\end{equation}

More detailed discussion of these transformations can be found in Sec. 10 of 
\cite{18}. (However, note a typo in the last term in eq. (84) there.)

Thus we can speak about the generalized Lema\^{\i}tre time applicable to
rotating systems (or simply Lema\^{\i}tre for shortness). We will use this
term even in a more complicated situation when a system is described by a
more general metric (\ref{mum}). These details do not affect our main
conclusions. Moreover, we will see below that our consideration applies also
to the Kerr metric with the role of $\theta $ coordinate taken into account.

\subsection{Near-horizon behavior of time and sign of $X$}

It follows from (\ref{t2}) that%
\begin{equation}
\bar{t}=\int_{r_{0}}^{r}(\frac{\sigma X}{PN\sqrt{A}}+\frac{z}{\Delta }%
)dr^{\prime }=-\int_{r}^{r_{0}}(\frac{X\rho \sigma }{P\sqrt{\alpha }}+z)%
\frac{dr^{\prime }}{\Delta }.  \label{t}
\end{equation}

Let a particle move outside the horizon towars a black hole with $r<r_{0}$.
Then, $\sigma =-1$, so%
\begin{equation}
\bar{t}=\int_{r}^{r_{0}}(\frac{X\rho }{P\sqrt{\alpha }}-z)\frac{dr^{\prime }%
}{\Delta }.
\end{equation}

When $r\rightarrow r_{+}$, $P\rightarrow \left\vert X\right\vert $. Outside
the horizon, using (\ref{forw}), we see that $P\rightarrow +X$. Then, the
main divergences in (\ref{t}) cancel due to (\ref{eq}) and $\bar{t}$ remains
finite.

In a similar manner, we can consider what happens inside a black hole.
Inside the horizon the variable $t$ and $r$ interchange their role, so $t$
becomes a spatial coordinate and $r$ becomes a time-like one. Therefore, eq.
(\ref{forw}) does not have the meaning of the forward-in-time condition and
is no-longer mandatory.

If a particle has $X>0$ and crosses the horizon, it is irrelevant whether we
consider time $\bar{t}$ from $r_{1}>r_{+}$ to $r_{+}$ or from $r_{+}$ to $%
r_{2}<r_{+}$. Anyway, it remains finite. However, the situation changes
radically, if $X<0$. In this case a particle cannot enter the inner region
from the outside since in the outer region negative $X$ is forbidden
according to (\ref{forw}). Meanwhile, it can travel there if it emerges
from, say, "mirror" universe or it appeared there as a result of particle
decay, etc. (but the mirror universe is beyond our consideration). Inside
the horizon $X<0$ is possible, nothing prevents it. To understand this
better, it is instructive to consider the metric there with the change of
variables $r=-T$, $t=y$ (see, e.g. page 25 of \cite{fn}) made directly in (%
\ref{met}). The sign is chosen so, that $r$ is decreasing when time $T$ is
passing, so we deal with particle motion inside a black (not white) hole.

Then, 
\begin{equation}
ds^{2}=-\frac{dT^{2}}{\left\vert \Delta \right\vert }\rho
^{2}+dy^{2}g+g_{\phi }(d\phi -\omega dy)^{2},  \label{under}
\end{equation}%
where $\Delta <0$ under the horizon and formally $N^{2}\rightarrow -g$, $%
g\geq 0$.

The equations of motion read%
\begin{equation}
m\frac{dT}{d\tau }=\frac{Z}{\rho \sqrt{\alpha }}\text{,}  \label{T}
\end{equation}%
\begin{equation}
m\frac{dy}{d\tau }=-\frac{X}{g},  \label{y}
\end{equation}%
\begin{equation}
Z=\sqrt{X^{2}+g(m^{2}+\frac{L^{2}}{g_{\phi }})}.  \label{Z}
\end{equation}

Thus automatically $\frac{dT}{d\tau }>0$ for any sign of $X$. Now, the
quantity $\sigma $ does not appear explicitly in eq. (\ref{y}) where $X\geq
0 $ or $X<0$ are admissible. A particle can move in any direction along the
leg of a hypercylinder since $X$ can have any sign.

As a result,%
\begin{equation}
\bar{t}=\int_{r}^{r_{0}}dr^{\prime }(\frac{z}{\left\vert \Delta \right\vert }%
-\frac{X}{gZ}\frac{\rho }{\sqrt{\alpha }})=\int_{-T_{0}}^{-T}dT^{\prime }(%
\frac{z}{\left\vert \Delta \right\vert }-\frac{X}{gZ}\frac{\rho }{\sqrt{%
\alpha }}).  \label{tx}
\end{equation}%
For $X>0\,\ $divergences cancel out mutually near the horizon. Now, let us
consider what happens inside the horizon when $X<0$. Here, there are two
cases to be examined.

The time from the past horizon $r=r_{+}$ to some point $r_{0}<r_{+}$ inside.
Then, it follows from (\ref{tx}) that $\bar{t}\rightarrow -\infty $.

The time from some point $r_{0}$ towards the inner horizon $r=r_{-}<r_{0}$.
Then, (\ref{tx}) with $X<0$ gives us that $\bar{t}\rightarrow +\infty $.

Thus in both cases (for the past or future horizons) the Lema\^{\i}tre time
diverges.

\subsection{Kerr metric: nonequatorial motion}

For the Kerr space - time the above results can be generalized to
non-equatorial trajectories since due to existence of the 3-d integral of
motion (the Carter constant) the variables in equations of motion can be
separated.

Indeed, the equations of motion in the Kerr space-time in the
Boyer-Lindquist coordinates are 
\begin{equation}
m\frac{dt}{d\tau }=-\frac{r_{g}ra}{\rho ^{2}\Delta }L+\frac{E}{\Delta }%
(r^{2}+a^{2}+\frac{r_{g}ra^{2}}{\rho ^{2}}\sin ^{2}{\theta })
\end{equation}%
and 
\begin{equation}
m^{2}(\frac{dr}{d\tau })^{2}=\frac{1}{\rho ^{4}}[(r^{2}+a^{2})E-aL]^{2}-%
\frac{\Delta }{\rho ^{4}}(K+m^{2}r^{2})
\end{equation}%
where, as usual, $\rho ^{2}=r^{2}+a^{2}\cos ^{2}{\theta }$ and $\Delta
=r^{2}-r_{g}r+a^{2}$ and $K$ is the Carter constant, $r_{g}=2M$, $M$ being a
black hole mass (see. e.g. \cite{LL}, page 329). The horizon corresponds to $%
\Delta =0$, where $r=r_{+}$ on the outer horizon and $r=r_{-}$ on the inner
horizon, $r_{\pm }=M\pm \sqrt{M^{2}-a^{2}}$, it is implied that $M>a$.

We can see that the influence of the Carter constant $K$ vanishes near the
horizon, where 
\begin{equation}
m\frac{dt}{d\tau }\approx \frac{(r_{+}^{2}+a^{2})}{\rho ^{2}(r_{+})\Delta }%
[E(r_{+}^{2}+a^{2})-aL]  \label{k1}
\end{equation}%
and 
\begin{equation}
m\frac{dr}{d\tau }=\frac{\sigma }{\rho ^{2}(r_{+})}[E(r_{+}^{2}+a^{2})-aL]%
\text{,}  \label{k2}
\end{equation}%
$\sigma =\pm 1$, whence%
\begin{equation}
\frac{dt}{dr}\approx \sigma \frac{(r_{+}^{2}+a^{2})}{\Delta }.
\end{equation}

One can introduce the Doran-Natario time $t^{\prime }$ \cite{doran}, \cite%
{natario} near the horizon according to $dt^{\prime }=dt+\frac{\sqrt{%
r_{g}r(r^{2}+a^{2})}}{\Delta }dr.$ (In the Schwarzschild case $a=0$ it
corresponds to eq. (102.1) of \cite{LL}.) Then, the finiteness or divergence
of this time at the horizon (due to the term $1/\Delta $) is determined by
the sign of the combination $(r^{2}+a^{2})E-aL$ at the horizon.

Meanwhile, for the Kerr metric 
\begin{equation}
\omega _{H}=\frac{a}{r_{+}^{2}+a^{2}}\text{,}
\end{equation}%
where $\omega _{H}=\omega (r_{+})$ has the meaning of the angular velocity
of the black hole. Therefore, near the horizon, the aforementioned
combination is proportional to $X$ given by eq. (\ref{x}), so it is the sign
of $X$ which is crucial in accordance with what is said in Sec. \ref{dirty}.

\section{Reissner-Nordstr\"{o}m black hole}

The same properties of the Lema\^{\i}tre time are valid also, if instead of
a rotating black hole we take the static charged one. Let us consider the
Reissner-Nordstr\"{o}m metric

\begin{equation}
ds^{2}=-fdt^{2}+\frac{dr^{2}}{f}+r^{2}d\omega ^{2}\text{, }d\omega
^{2}=d\theta ^{2}+\sin ^{2}\theta d\phi ^{2}\text{,}
\end{equation}%
\begin{equation}
f=1-\frac{2M}{r}+\frac{Q^{2}}{r^{2}}\text{,}
\end{equation}%
the Coloumb potential%
\begin{equation}
\varphi =\frac{Q}{r}\text{.}
\end{equation}

Here, $M$ is a black hole mass, $Q$ being its electric charge. We assume $%
Q>0 $. For simplicity, we consider pure radial motion. Now, the event
horizon is located at $r_{+}=M+\sqrt{M^{2}-Q^{2}}$, $M>Q$. The inner horizon
is located at $r_{-}=M-\sqrt{M^{2}-Q^{2}}$.

Then, we take advantage of the approach developed in \cite{radobs}. One can
introduce the Lema\^{\i}tre time $\tilde{t}$ according to%
\begin{equation}
dt=\frac{1}{e_{0}}(d\tilde{t}-\frac{dr}{f}p_{0})\text{,}
\end{equation}%
where $e_{0}$ is the specific energy of a fiducial observer whose set
compose the frame, \thinspace $p_{0}=\sqrt{e_{0}^{2}-f}$. It is implied that
corresponding particles are electrically neutral.

One reservation is in order. For the standard Lema\^{\i}tre frame, fiducial
observers that compose this frame, fall from infinity with zero velocity. In
this subsection, there is no such a restriction. The free falling frame can
be formed by particles having non-zero velocity at infinity (this case
corresponds to the parameter $e_{0}$ being bigger than $1$) or falling from
a rest at some finite $r$ (for this case $e_{0}<1$). The crucial point under
discussion concerns behavior near the horizon, while what happens far from
it is irrelevant. In this sense, our treatment is more general that in the
standard case of the Lema\^{\i}tre frame which corresponds to $e_{0}=1$.
Since our qualitative results appear to be independent of $e_{0}$ we
continue to refer to this generalized free falling frame time as the Lema%
\^{\i}tre time.

The metric now reads%
\begin{equation}
ds^{2}=-d\tilde{t}^{2}+(dr+\frac{p_{0}}{e_{0}}d\tilde{t})^{2}+r^{2}d\omega
^{2}\text{.}
\end{equation}

If, additionally, we introduce a new variable $\chi $ according to 
\begin{equation}
d\chi =\frac{dr}{p_{0}}+d\tilde{t}\text{,}
\end{equation}%
we obtain the standard Lema\^{\i}tre form%
\begin{equation}
ds^{2}=-d\tilde{t}^{2}+\frac{p_{0}^{2}}{e_{0}^{2}}d\chi ^{2}+r^{2}(\chi ,%
\tilde{t})d\omega ^{2}\text{.}  \label{RL}
\end{equation}

For radial fall of a particle with the specific energy $e$ and electric
charge $q$ we have%
\begin{equation}
m\frac{dt}{d\tau }=\frac{X}{f}\text{,}
\end{equation}%
\begin{equation}
X=E-q\varphi \text{,}  \label{xrn}
\end{equation}%
\begin{equation}
m\frac{dr}{d\tau }=\sigma P\text{,}
\end{equation}%
\begin{equation}
P=\sqrt{X^{2}-m^{2}f}\text{.}
\end{equation}

These equations for a charged particle differ from those for a neutral one
by the replacement $e\rightarrow \frac{X}{m}$. We would like to stress that,
by contrast with variable $t$, the Lema\^{\i}tre time $\tilde{t}$ retains
its time-like character both outside and inside the horizon,as this is seen
from (\ref{RL}).

Then,%
\begin{equation}
\frac{dr}{d\tilde{t}}=\frac{Pf}{Pp_{0}+\sigma Xe_{0}},
\end{equation}%
\begin{equation}
\tilde{t}=\int^{r}\frac{dr^{\prime }}{Pf}(Pp_{0}+\sigma Xe_{0}).
\end{equation}

\subsection{R-region}

Let a particle move outside the horizon (in the R-region, according to
classification \cite{novrt}) towards the horizon, $\sigma =-1$, $r_{0}>r$.
According to the forward-in-time condition, $X>0$. We have%
\begin{equation}
\tilde{t}=\int_{r}^{r_{0}}\frac{dr^{\prime }}{Pf}(Xe_{0}-Pp_{0})\text{.}
\end{equation}

When $r\rightarrow r_{+}$ then $P\rightarrow X$, $p_{0}\rightarrow e_{0}$,
the numerator has the order $f$ and compensates the denominator, so $\tilde{t%
}$ is finite.

However, if $\sigma =+1$, $\tilde{t}$ diverges. Note that since the particle
moves now not to a horizon but from a horizon, it is the time elapsed from
the \textit{past} horizon to current position of the particle which is
infinite.

\subsection{T-region}

Inside the horizon, in the black hole region, $f=-g$, $r=-T$, $t=y$ and%
\begin{equation}
m\frac{dT}{d\tau }=Z\text{,}  \label{z1}
\end{equation}%
\begin{equation}
Z=\sqrt{X^{2}+m^{2}g}\text{,}
\end{equation}%
\begin{equation}
m\frac{dy}{d\tau }=-\frac{X}{g}\text{,}  \label{y1}
\end{equation}%
\begin{equation}
\tilde{t}=\int_{T_{0}}^{T}\frac{dT^{\prime }}{Pg}(\frac{ZZ_{0}}{mm_{0}}%
-Xe_{0}),
\end{equation}%
where $Z_{0}=m_{0}\sqrt{e_{0}^{2}+g}$.

If $X>0$, $\tilde{t}$ between the event horizon and a point inside the $T$%
-region is still finite as well as $\tilde{t}$ between such a point and the
inner horizon. However, inside a black hole $X<0$ is also possible for the
same reasons as was described in Sec. \ref{II}. Then, $\tilde{t}$ diverges
logarithmically in the vicinity of the horizon.

The above results are valid also for the Schwarzschild metric ($Q=0$, $X=me$%
). Note, however, that since $e<0$ is possible under the horizon only, where 
$e$ has a physical meaning of momentum (not energy) due to interchange
between temporal and spatial coordinates, the motion toward a horizon is
impossible in the $T$-region (i.e. inside a black hole). This means that for
the Schwarzschild black hole the divergent time is the time from the \textit{%
past} horizon for both $R$-region and $T$-region cases. Motion to a future
horizon is possible only in the $R$-region and is directed inward, so the
Lema\^{\i}tre time to a future horizon of a Schwarzschild black hole is
always finite.

Combining the results of this section with our white hole results of \cite%
{white-black} it is worth making a general remark for the Schwarzschild
space-time. In general, for eternal black-white hole and radially moving
observers there exist four different Lema\^{\i}tre frames. If we consider an
astrophysical black hole and thus discard the mirror universe, only two of
them remain relevant - contracting and expanding ones. And, divergence of
the Lema\^{\i}tre time signals that we try to describe an inward motion in
the expanding frame, more adjusted for description of the outward motion (or
vice versa). In other words, this is manifestation of incompleteness of the
frame under discussion, and a particle is just near the boundary beyond
which a corresponding frame fails.

\subsection{Properties of X in particle collision and its meaning}

The sign of quantity $X$ which is so crucial, not only determines the
behavior of the Lema\^{\i}tre time, it is responsible for possibility of
high energy collisions. Let two particles 1 and 2 collide under the horizon.
The energy in the center of mass frame is defined according to $%
E_{c.m.}^{2}=-P_{\mu }P^{\mu }$, where $P^{\mu }=m_{1}u_{1}^{\mu
}+m_{2}u_{2}^{\mu }$. Then, 
\begin{equation}
E_{c.m.}^{2}=m_{1}^{2}+m_{2}^{2}+2m_{1}m_{2}\gamma \text{,}
\end{equation}%
where $\gamma =-u_{1\mu }u^{2\mu }$ is the Lorentz gamma factor of relative
motion.

It follows from equations of motion \ (\ref{T}) - (\ref{Z}) or (\ref{z1}) - (%
\ref{y1}) that%
\begin{equation}
m_{1}m_{2}\gamma =\frac{Z_{1}Z_{2}-X_{1}X_{2}}{g}\text{.}  \label{ga}
\end{equation}%
In the simplest case of the Schwarschild metric and equal masses (\ref{ga})
reduces to eq. (8) \ of \cite{inner}.

Under the horizon, $X$ can be negative. Then, if $X_{1}X_{2}<0$ (say, $%
X_{1}<0$ and $X_{2}>0$)$,$ the quantity $E_{c.m.}$ becomes as large as one
like when a point of collision approaches the horizon, so $g\rightarrow 0$.
But it cannot be literally infinite because two particles 1 and 2 do not
meet in the same point. The above result just explains this fact using the
language of the Lema\^{\i}tre time: in the horizon limit $\bar{t}_{1}$
diverges and $\bar{t}_{2}$ remains finite.

In previous consideration, the quantity $X$ (\ref{x}) was written in the
particular coordinate system (\ref{met}).\ Meanwhile, it can be presented in
a coordinate-independent form. The energy of a particle $E=-mu_{\mu }\xi
^{\mu }$ where $\xi ^{\mu }$ is the Killing vector responsible for
translations along $t$ and $u_{\mu }$ is the four-velocity. The metric
coefficient $\omega =-\frac{g_{0\phi }}{g_{\phi }}$. Here, $g_{0\phi
}=g_{\mu \nu }\xi ^{\mu }\eta ^{\nu }$ and $g_{\phi }=g_{\mu \nu }\xi ^{\mu
}\eta ^{\nu }$, where $\eta ^{\mu }$ is the Killing vector responsible for
rotations along the polar axis. As a result,%
\begin{equation}
\frac{X}{m}=-u_{\mu }\xi ^{\mu }+\frac{g_{\mu \nu }\xi ^{\mu }\eta ^{\nu }}{%
g_{\mu \nu }\eta ^{\mu }\eta ^{\nu }}u_{\alpha }\eta ^{\alpha }\text{.}
\label{X}
\end{equation}%
Under the \ horizon, the vector $\xi ^{\mu }$ changes its character and
becomes space-like, but general formula (\ref{X}) remains valid.

In the electromagnetic case 
\begin{equation}
P_{\mu }=\mathcal{P}_{\mu }-qA_{\mu }\text{,}
\end{equation}%
where $P_{\mu }=mu_{\mu }$ is the kinematic momentum, $\mathcal{P}_{\mu }$
is the generalized one, $A_{\mu }$ is the vector potential. For the
Reissner-Nordstr\"{o}m metric $A_{\mu }=(-\varphi $, $0$, $0$, $0)$. Then, $%
X=-P_{\mu }\xi ^{\mu }$, the Killing energy being $E=-\mathcal{P}_{\mu }\xi
^{\mu }$. Thus $X$ corresponds to the kinematic momentum, it is $E~$but not $%
X$ which is conserved.

In the rotating case, $\omega $ is the analogue of the potential and $L$ is
the analogue of the electric charge. In this sense, eq. (\ref{x}) is similar
to the expression (\ref{xrn}) that relates the kinematic and generalized
momenta.

Although we did not consider the mirror universe, it is instructive to make
an important methodical remark about it. We would like to stress that
property $X<0$ under discussion occurs between horizons and is of dynamic
origin, being depending on the relation between the energy and electric
potential. This should not be confused with seemingly similar property $X<0$
in the "mirror Universe". In such a universe (which lies beyond the inner
horizon) the variable $t$ runs into the past, so $X<0$ arises just due to
causality requirement. There, it is consequence of kinematics and is valid
for any relation between the energy and potential (or energy and angular
momentum in the rotating case). In other words, there are two qualitative
different cases for different regions of space-time in which $X<0$.

%We would also like to stress that property $X<0$ under discussion occurs
%between horizons and is of dynamic origin, being depending on the relation
%between the energy and electric potential. This should not be confused with
%seemingly similar property $X<0$ in the "mirror Universe". In such a
%universe (which lies beyond the inner horizon) the variable $t$ runs into
%the past, so $X<0$ arises just due to causality requirement. There, it is
%consequence of kinematics and is valid for any relation between the energy
%and potential (or energy and angular momentum in the rotating case). In
%other words, there are two qualitative different cases for different regions
%of space-time in which $X<0$.

\section{Conclusions}

In our recent paper \cite{white-black} we considered collision of two
particles 1 and 2 in the Schwarzschild background in the $R$ region near the
horizon. In doing so, particle 1 moved entirely in the $R$ region whereas
particle 2 emerged from the $T^{+}$ one corresponding to the white hole. We
analyzed there two separate scenarios for collisions near a white and black
hole horizons. In both cases particles had Killing energies $E_{1,2}>0$ but
their radial momenta had different sign: $\dot{r}_{1}=-\left\vert \dot{r}%
_{1}\right\vert $, $\dot{r}_{2}=+\left\vert \dot{r}_{2}\right\vert $. It
turned out that high energy collision is possible but $E_{c.m.},$ however
big it be, remains finite. One can try to arrange collision with literally
infinite $E_{c.m.}$ but this requires collision exactly on the horizon. This
leads to a situation when in the free falling frame (outgoing Lema\^{\i}tre
frame) time of the particle with a positive $\dot{r}_{1}$ crossing the white
hole horizon is finite while the same time for the particle with a negative $%
\dot{r}_{2}$ is infinite. This makes the collision between such particles
impossible, avoiding physically unacceptable situation of an infinite
collision energy $E_{c.m.}$, so kinematic censorship \cite{cens} is
preserved. Since for the consideration in \cite{white-black} the particular
form of the metric has not been used, the same arguments are applicable to
collisions of two neutral particles near the outer horizon of the
Reissner-Nordstr\"{o}m black holes.

The present paper generalizes these results in three ways. First, we
considered the motion of charged particles in the Reissner-Nordstr\"{o}m
black hole background whose trajectories are not geodesics. Second, we
considered a general class of axially-symmetric metrics and obtained the
results for equatorial motion. \ For the particular but physically important
case of the Kerr metric we obtained the results for an arbitrary geodesic
motion. Third, now we considered particle motion including collision inside
the horizon. It turns out that in all these cases there exists a quantity $X$
which for collisions inside the black hole horizon plays a role similar to
the radial momentum for collisions outside it \cite{white-black}. This
quantity has the following basic properties relevant in our context:

\begin{itemize}
\item $X<0$ can take place only in the $T$-region (provided we do not
consider the mirror Universe).

\item The free falling frame time (generalized Lema\^{\i}tre time for a
spherically symmetric black hole or the generalized Doran - Natario time for
an axially symmetric black hole) needed for a particle with $X<0$ to reach
the horizon is infinite, while for particle $\ $with $X>0$ it is finite.

\item Collision energy $E_{c.m.}$ of two particles with opposite signs of $X$
colliding exactly at a horizon is infinite.
\end{itemize}

Combining the 2-nd and 3-d properties we see that collisions giving infinite
energy are physically unrealizable, that generalizes \cite{white-black} .

Thus we established intimate connection between different phenomena that
occur near the horizon. This includes the behavior of Lema\^{\i}tre time,
high energy particle collisions and validity of the kinematic censorship.
All these aspects are unified by the properties of the quantity $X$ and,
especially, its sign. These properties and their interrelations are valid
for rotating and non-rotating black and white holes. In particular, this
includes the Reissner-Nordstr\"{o}m and even Schwarzschild (where $X$
reduces to Killing energy/momentum $E$) ones. Thus we gave a unified picture
of what seemed to be separate issues.

Now, our results obtained in the present paper and the previous one \cite%
{white-black} encompass two situations: particle collision near the inner
horizon of \ black hole (when $X_{1}$ and $X_{2}$ have different signs) and
near the outer horizon of a white hole (where radial velocities $\dot{r}_{1}$
and $\dot{r}_{2}$ have different signs).

Apart from the context connected with particle collisions, general results
concerning the behavior of the Lema\^{\i}tre time for an individual
particle, can be of some use for general analysis of particle trajectories
in black hole background.

\section{Acknowledgement}

The authors thank Te\'{o}filo Vargas for discussions, and Grupo de Astronom%
\'{\i}a SPACE and Grupo de F\'{\i}sica Te\'{o}rica GFT, Facultad de Ciencias
F\'{\i}sicas, Universidad Nacional Mayor de San Marcos (Lima, Per\'{u)}),
where part of this work have been done, for hospitality. OZ was supported in
part by the grants 2024/22940-0, 2021/10128-0 of S\~{a}o Paulo Research
Foundation (FAPESP).

\end{document}